\documentstyle[titlepage,12pt,epsfig]{article}

\def\be{\begin{equation}}

\def\ee{\end{equation}}

\def\to{\rightarrow}

\newcount\sectionnumber
\def\sect
{\global\equationnumber=0
\global\advance\sectionnumber by 1
\the\sectionnumber . }
\newcount\equationnumber
\def   \num
{\eqno{\global\advance\equationnumber by 1
\left(\the\sectionnumber .\the\equationnumber \right)}}
\begin{document}
\title{Formation of Topological Black Holes from Gravitational Collapse}
\author{W.L. Smith and R.B. Mann\\
        Department of Physics\\
        University of Waterloo\\
        Waterloo, Ontario\\
        N2L 3G1}
\maketitle
\begin{abstract}
We consider the gravitational collapse of a dust cloud in an asymptotically 
anti de Sitter spacetime in which points connected by a discrete subgroup
of an isometry subgroup of anti de Sitter spacetime are identified.
We find that black holes with event horizons of any topology can form from
the collapse of such a cloud. The quasilocal mass parameter of such black holes
is proportional to the initial density, which can be arbitrarily small.
\end{abstract}

\section{Introduction}

Black hole solutions to the Einstein field equations play a substantial role in our 
understanding of gravitation.  They provide us with an important arena for testing 
some of our most fundamental ideas about thermodynamics and quantum physics.  An increasingly
large body of evidence that black holes do indeed exist in our universe as physical entities
\cite{physblack}, and are not merely mathematical constructs, motivates the study of 
such objects even further.

There has been growing interest in recent years concerning more exotic black-hole-type
solutions to the Einstein field equations, particularly those with non-trivial topology.
Although it was postulated by Freidmann {\it et al.} \cite{wita} that in a globally
hyperbolic, asymptotically flat spacetime topologically non-trivial event-horizon
structures cannot be observed (they collapse before light can traverse them), it was later
shown that such topologies could be passively observed \cite{witb}.  In addition to this,
it has been shown numerically that black hole event horizons with toroidal
topology can form, at least temporarily \cite{Shapiro}.
More recently Aminneborg {\it et. al.} have shown that by suitably identifying
points in $(3+1)$-dimensional anti de Sitter (AdS) spacetime, black hole solutions can result
\cite{amin}.
The black holes they constructed were eternal black holes whose event horizons had non-trivial
topology of genus higher than one. The construction of these AdS black holes are
generalizations of the construction in $(2+1)$ dimensions of Ba\~{n}ados, Teitelboim and
Zanelli (BTZ) \cite{bana}, and the
resultant  $(3+1)$-dimensional black holes can be considered higher dimensional analogues of
the lower-dimensional case.

It is natural to ask to what extent this type of black hole can arise from known, or at least
hypothetically plausible, physical processes.  It was recently shown that pair-production
of such black holes of arbitrary genus is possible in the presence of a domain wall of
suitable topology. In addition to 
this, solutions with non-zero quasi-local mass and
charge were obtained \cite{adscmet}, generalizating the constructions given in
ref. \cite{amin} to spacetimes with non-constant curvature.

We investigate here the possibility of forming this type of black hole from the collapse
of a cloud of pressureless, uncharged dust.  Traditionally, the study of gravitational collapse into
black holes has been limited to spheres of collapsing dust and gas.  We consider
a cloud of dust embedded into a region created by identifying points on a surface of
constant negative curvature.  This is a generalization of the collapse process considered
in $(2+1)$ dimensions \cite{ross}, in which it was shown that a disk of pressureless
dust could collapse into a BTZ black hole provided the initial density was sufficiently
large. Although we find that black holes of non-trivial topology can form from gravitational
collapse in a non-asymptotically flat spacetime, we do not find an analogous constraint on
the initial density; rather we find that arbitrarily small initial densities of pressureless
dust will collapse to form such black holes.

The outline of our paper is as follows. We consider in section II the general structure of
the exterior and interior metrics.  In section III we solve for the metric inside the dust
cloud, and in section IV we match this solution to the exterior metric.  We consider
the case of collapse to a `massless black hole' in section V, and summarize our results
in section VI.

\section{Topological Black Hole Metrics}

        A further generalization of the black hole metrics discussed in ref. \cite{amin}
was recently obtained in the context of investigating which cosmological C-metrics
could provide suitable instantons for black hole pair-production \cite{adscmet}.
These metrics included terms for a nonzero mass and charge, thereby yielding  spacetimes
of varying curvature.  We consider here collapse to neutral black holes only, for which
the metric outside of the dust reads
\be
ds^2=-\left( \frac{\Lambda}{3}R^2+b-\frac {2M}{R}\right) dT^2
+\frac{dR^2}{(\frac{\Lambda}{3}R^2+b- \frac{2M}{R})}
+R^2 \left( d\hat \theta ^2 +  c \sinh ^2 (\sqrt d \hat \theta ) d \hat \phi ^2 \right),
\label{eq1}
\ee
where $T$ is the time coordinate, $R$ is the radial coordinate, and $\hat \theta$ and
$\hat \phi$ are coordinates on a 2-surface of constant curvature, $\hat{\phi}$ being an angular
coordinate whose range is from $0$ to $2\pi$. $\Lambda$ refers to the cosmological constant,
with positive $\Lambda$ corresponding to anti de Sitter spacetime, and $M$ is a
constant of integration corresponding to the quasi-local mass \cite{adscmet}.

Solving the Einstein field equations for empty space, it is clear that $b=-d$ is the only solution.
Without loss of generality we may set the magnitudes of $b$ and $d$ to unity (or zero). There
are then three possibilities \cite{adscmet}:
\vskip .25in
{
$1$.  If { }$b=-d=+1$, c is forced to be $-1$ in order to preserve the
signature of the metric.  This is suitable for spaces with positive, negative or zero
asymptotic curvature.  Surfaces of constant $(R,T)$ are two-spheres, and the
topology of spacetime is ${\bf R}^2\otimes {\bf S}^2$.
The cosmological constant may be positive or negative.  The resultant spacetime is
simply Schwarzchild (anti) de Sitter space.

$2$.  If\/ $b=-d=0$, the signature forces $c=+1 /d$ { }in the limit $d \to 0$.
$\Lambda$ in this case is  strictly positive.  The metric corresponds to a spacetime
that is asymptotically anti de Sitter
in the $(R,T)$ section.  The topology is that of a torus ${\bf R}^2\otimes {\bf T}^2$.

$3$.  If\/ $b=-d=-1$, then $c=+1$ and $\Lambda$ is again strictly positive.
In this case the topology is ${\bf R}^2\otimes {\bf H}_g^2$, where ${\bf H}_g^2$
is a 2-surface of constant negative curvature and genus $g > 1$.}

\vskip .3in

In case $2$, without suitable identifications in the $(\hat{\theta},\hat{\phi})$ sector,
the black hole event horizon would be an infinite sheet.  Such `black plane' solutions
have been recently noted in the literature \cite{bplane}. However, a black hole with a
compact event horizon may be obtained by considering
only a portion of the sheet, and making this section periodic through the
suitable identification of points.  The shape for the section chosen
is a polygon formed from geodesics in the $(\hat{\theta},\hat{\phi})$ sector, in this
case straight lines, cut out of the plane.  The sum of the angles in the polygon must be equal
to $2 \pi $ or greater in order to avoid the formation of conical singularities.  In addition,
the number of sides must be an integer multiple of four \cite{amin}.  The simplest polygon
fulfilling these criteria is a square.  Opposite sides of the square are identified,
yielding the toroidal toplogy noted above.

Turning now to case 3, which has the exterior metric
\be \label{exthyp}
ds^2=-\left({\Lambda \over 3}R^2-1-{2M \over R}\right) dT^2 +{dR^2 \over {\Lambda \over 3} R^2-1-{2M
\over R}}
+R^2(d \hat \theta ^2 +\sinh^2 \hat \theta d \hat \phi^2),
\ee
where the $(\hat{\theta},\hat{\phi})$ sector is now a
space with constant negative curvature, also known as a hyperbolic plane  or
a pseudosphere. These spaces have been discussed in detail by Balazs and Voros \cite{bala}.
Geodesics on the pseudosphere are formed  from intersections of the
psuedosphere with planes through the origin, and are the analogs of
great circles on  a surface of
constant positive curvature (a sphere), which are intersections of
the sphere and planes through the origin.
A projection of the psuedosphere onto the $(y_1, y_2)$ plane
is known as a the Poincar\'e disk. On it, geodesics
are segments of circles, orthogonal to the disk boundary at the edges
\cite{bala}.  The pseudosphere, its associated
Poincar\'e disk and the geodesics are shown in figure 1.
  \begin{figure}
    \leavevmode
    \hfil \hbox{\epsfysize=7cm \epsffile{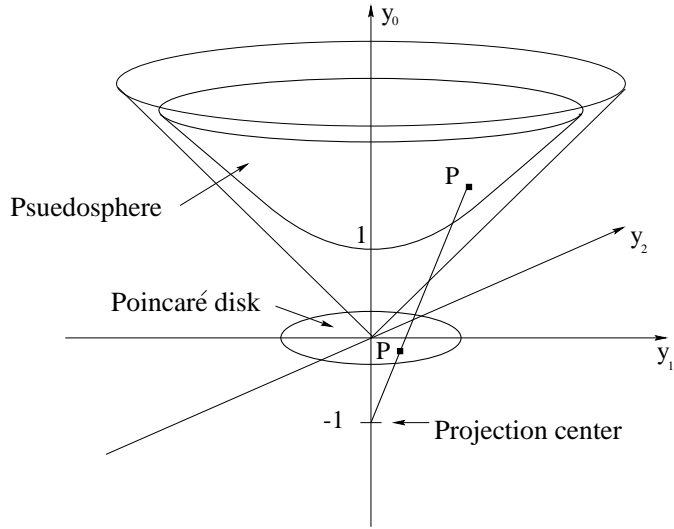}} \hfil
    \caption{The pseudosphere is one half of the hyperboloid.  The axes do not represent
any coordinates in particular.  Beneath the pseudosphere is the Poincar\'e disk, the center of which
is the origin. [Balazs and Voros, p121]}
    \label{F1}
  \end{figure}

A compact surface on the pseudosphere can be obtained by identifying opposite sides of
a suitably chosen polygon centered at the origin. In order to avoid conical singularities,
we must construct a polygon from geodesics which has angles that sum to
$2 \pi$ or more, and a number of sides that is a multiple of four.
Since the geodesics on the pseudosphere meet at angles smaller than those for geodesics
meeting on a flat plane, an octagon is the simplest solution, yielding a surface of
genus 2. This construction is shown in figure 2.  In general, a polygon of $4g$ sides
yields a surface of genus $g$, where $g\geq 2$.
  \begin{figure}
    \leavevmode
    \hfil \hbox{\epsfysize=10cm \epsffile{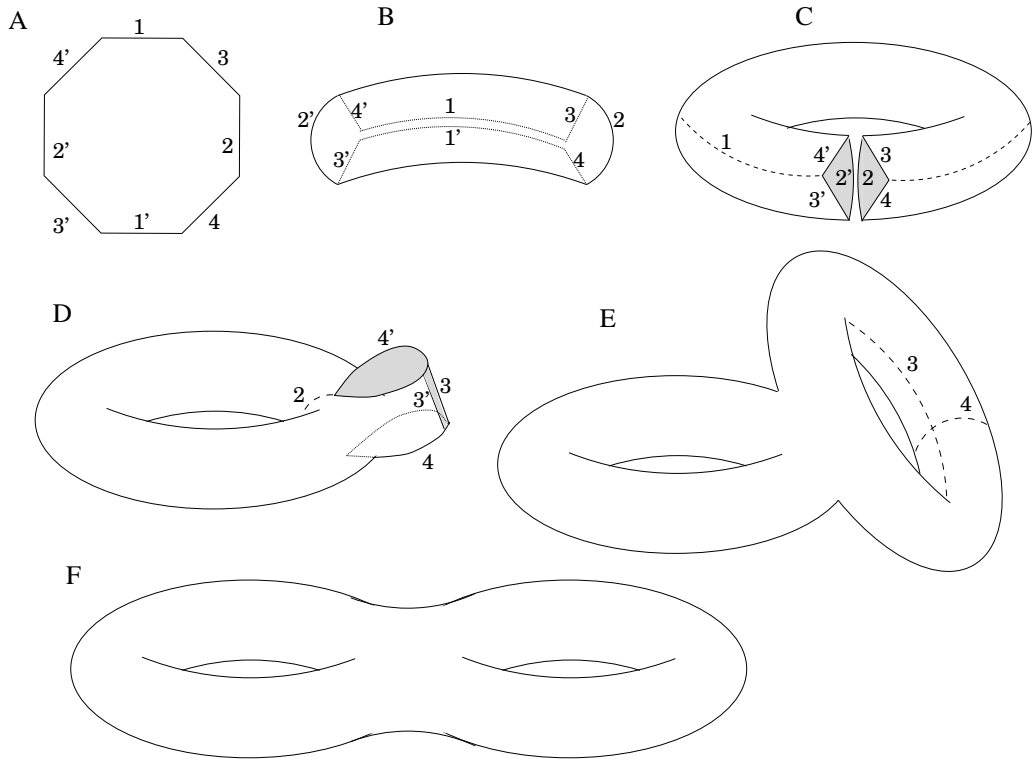}} \hfil
    \caption{The identification of the octagon is shown.  Opposite sides of the octagon 
in 3A will be identified.  The sides are drawn straight for clarity.  Dashed lines indcate where sides
have been sewn together.    First, sides $1$ and $ 1^\prime$ are identified, folding the 
top and  bottom of the octagon away from view (3B).  Sides $2$ and $2^\prime$ are 
brought together to  form a torus with a
diamond shaped  hole, as in 3C.  Next, sides $3$ and $3^\prime$ are stretched out and joined 
for 3D.  The loop is lengthened along the direction of identification 3 and bent until $4$ and 
$4^\prime$ meet, forming a second torus.  Finally the topology is deformed to the preferred 
shape, seen in 3F.  Identification of a polygon of genus $g$ will clearly result in $g$ attached tori
 or, equivalently, a $g$-holed pacifier.}
\label{F2}
 \end{figure}

We shall consider in this paper collapse of a cloud of dust to a black hole whose
exterior metric is given by (\ref{exthyp}), keeping in mind that our procedure easily
generalizes to the toroidal case.
In order for a black hole to form, two conditions must be satisfied.
First, the exterior metric must be matched successfully to a metric for the interior of the
dust cloud. Secondly, the dust cloud must be shown to collapse in a finite amount of
proper time.  Simple criteria of homogeneity and topology suggest that we
match the above spacetime metric (\ref{exthyp}) to an interior metric of the form
\be
ds^2=-dt^2+ a^2(t) \left(\frac{dr^2}{b- \tilde{k} r^2}+r^2 \left( d\theta^2
+ c \sinh^2({\sqrt{d}}\theta) d\phi^2) \right) \right),
\label{eq2}
\ee
where $r$, $\theta$ and $\phi$ are the comoving radial and
angular interior coordinates respectively. $t$ is the proper time
of the dust cloud and $a(t)$ is the scale factor.  $\tilde{k}$ refers to the  curvature.
Again, requiring the above metric to be an exact solution of the Einstein equations
with the stress energy tensor being that of pressureless dust yields $b=-d=-c=-1$ and
$\tilde{k} < 0$, so that the form of the interior metric becomes
\be\label{inthyp}
ds^2=-dt^2 + {a(t)^2 \over k r^2-1}dr^2 + r^2  a(t)^2 (d\theta^2+\sinh^2\theta d\phi^2).
\ee
where we have subsituted $k = |\tilde{k}|$.

In the next section, we shall solve Einstein's equations inside the dust cloud.

\section{The Interior Solution}

Consider a collapsing dust cloud, surrounded by a vacuum, where the metric for the dust is
given by (\ref{inthyp}).  It remains for us to solve for the scale factor $a(t)$,  checking
to see under what circumstances (if any) $a(t)$ vanishes in a finite amount of proper time,
$t$.

The stress energy tensor for freely-falling dust is
given by $T_{\mu\nu} = \rho u_\mu u_\nu$, where $u_\nu$ is the four velocity, defined such
that  $g^{\mu\nu} u_\mu u_\nu = -1$.  In comoving coordinates, this means
$u_\nu=(-1,0,0,0)$.  If the initial density is given by $\rho _0$ and the scale factor is
initially $a_0$, the conservation of stress energy, $T^{\mu \nu}{}_{; \nu}=0$, will
require that $ \rho (t) (a(t))^3 = \rho_0 a_0^3$. In addition, requiring
that the dust satisfy the standard positive energy criteria implies that $\rho(t) \geq 0$.

The relevant form of the Einstein field equations with a non-zero cosmological constant is given by
\be
G_{\mu\nu} = 8 \pi G T_{\mu\nu} + \Lambda g_{\mu\nu}.
\label{eq3}
\ee
Application of this yields two equations: from the temporal component, 
\be
\dot a^2 = {-\Lambda \over\ 3} a^2 + {8 \over 3} \pi G {a^3_0 \rho _0 \over a} + k ,
\label{eq4}
\ee
and from the three spatial components,
\be
2 a \ddot a + \dot a^2 - k + \Lambda a^2=0,
\label{eq5}
\ee
where the overdot is used to denote the time derivative, $d/dt$.  The second equation is directly 
derivable from the first, leaving a single independent equation.

{}From our earlier discussion we know that $k$ must be strictly greater than zero, implying

\be
\Lambda + 3\frac{\dot{a}^2_0}{a^2_0} > 8 \pi G \rho_0,
\label{eq6}
\ee
from (\ref{eq4}), where $\dot{a}_0$ is the initial inward velocity of the cloud.
This somewhat counterintuitive condition reflects the need to maintain negative curvature 
within the dust cloud.  
An initial density that is too large could reverse the sign of the curvature  
of the spatial sections of its spacetime.
The more familiar case 1, where $b=-d=+1$ and $c=-1$, has no analogous limits since the 
form of its metric allows a free choice in the sign of the curvature.  
No such choice is admissible for our metric in a manner that preserves the 
spacetime signature.

When the cosmological constant is absorbed into the
definition of the  stress-energy tensor of the Einstein equations, it becomes
\be
\tilde T_{\mu \nu} =\tilde p g_{\mu \nu} + (\tilde \rho +\tilde p)u_\mu u_\nu 
\ee
where 
\be
\tilde \rho = \rho - {\Lambda \over 8 \pi G} \qquad \tilde p = p + {\Lambda \over 8 \pi G}.
\ee
In other words, we can consider the cosmological constant as in some manner contributing to the
pressure in our space and lowering the density therein.  If $\Lambda$ is not of sufficient size,
or we do not give the cloud enough initial inward velocity, collapse cannot occur. 
In fact, for collapse from rest, we require that the net effective density from matter and the 
vaccum energy be negative.  If this condition is not satisfied, the metric will change 
its signature and will no longer properly describe the dust cloud.

Assuming the condition (\ref{eq6}) is satisfied, the dust will abide by the equation of 
motion given by (\ref{eq4}).  Solving (\ref{eq4}) yields
\be
d\tilde{t} = dx \sqrt{ x \over {-x^3+(1+v_0^2)\left((1-B)x+B \right)}},
\label{eq8}
\ee
where $\tilde{t} = \sqrt{\frac{\Lambda}{3}}$ $ t$, $x$ is the relative scale
factor, $a/a_0$, and $v_0 \equiv \frac{d x}{d\tilde{t}}\vert_{\tilde{t}=0}$. The
parameter $B$ is  given by
\be
B= {8 \pi G \rho_0 \over (1+v_0^2)\Lambda}
\label{eq8a}
\ee
and has the range $1>B>0$.
We now have a parametric solution for the scale factor, $a(t)$.  While it is
possible to generate an exact solution to this equation, it is irrelevant for our purposes.
It is sufficient for us to know that $a(t)$ exists as some combination of elliptic functions.

Since the interior metric is defined, we can move on to finding the collapse time.
If we start from time $t=0$, we can use (11) to find the time until complete collapse in the
interior coordinates.  The collapse time, $t_c$, is found by integrating (\ref{eq8}) from
$x=1$ to $x=0$:
\be
\tilde{t}_c = \int\limits_1^0 dx \sqrt{ x \over {-x^3+(1+v_0^2)\left((1-B)x+B \right)}}
\label{eq9}
\ee
This integral can be evaluated numerically. We find that
the collapse time is finite for all allowed $B$ between $0$ and $1$, and for
all allowed $v_0$.  The collapse time for a range of $v_0$ and all allowed  values of $B$ is shown in 
figure 3.

  \begin{figure}
    \leavevmode
    \hfil \hbox{\epsfysize=6cm \epsffile{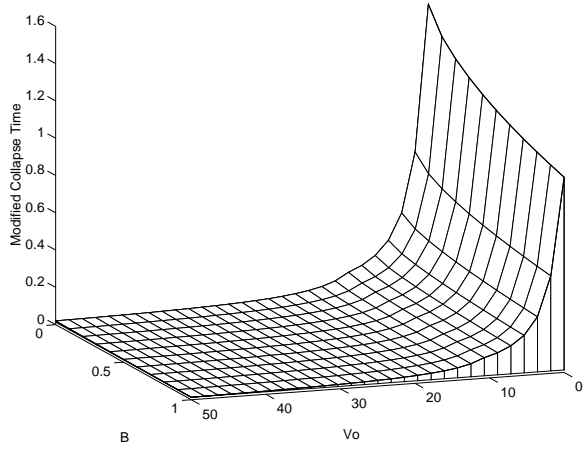}} \hfil
    \caption{The collapse time times ${\sqrt{\Lambda \over 3}}$ is shown as a function of $B$ and
$v_0$.  Increasing either $B$ or $v_0$ speeds collapse.}
    \label{F3}
  \end{figure}

\section{Matching Conditions}

In the exterior coordinates, the stress-energy vanishes and the exterior metric (\ref{exthyp})
provides the correct description of spacetime.  Our final condition on the collapse process
will be the requirement that the boundary between the two spacetimes be
smooth, with no energy shells separating them.
In order to make the dust edge a boundary surface with no shell of stress-energy present,
we will require that \cite{israel, chase}
\be
[g_{ij}]=0   \qquad\mbox{and}\qquad   [K_{ij}]=0      \label{eq13}
\ee
where $[\Psi]$ denotes the discontinuity in $\Psi$ across the edge, $K_{ij}$ is the extrinsic
curvature  of the dust edge, and the subscripts $i,j$ refer to the coordinates on the dust edge.

The metric on the edge of the dust cloud is
\be
ds^2 = - d \tau ^2 + \Re^2(\tau) (d \tilde \theta^2 + \sinh^2 \tilde \theta d \tilde \phi ^2),
\label{eq11}
\ee
where $\tau$, $\tilde \theta$ and $\tilde \phi$ are the boundary coordinates.
In the interior coordinates, $r=r_0$ at the
boundary. In the exterior coordinates, the boundary is at $R= \Re(\tau)$.
Continuity of the metric immediately implies  $ \theta=\hat \theta=\tilde \theta$ and
$ \phi=\hat \phi=\tilde \phi$. It
is also possible to choose $\tau = t$.  The boundary metric then becomes
\be
ds^2 = - d t^2 + \Re^2(t) (d \theta^2 + \sinh^2 \theta d \phi ^2).
\label{eq12}
\ee
Successful matching of the interior and exterior metrics forces
\be
\left({\Lambda \over 3} \Re^2 - 1 -{{2M} \over {\Re}}\right) \dot T^2-
{\dot \Re^2 \over({\Lambda \over 3}  \Re^2 -1 -{2M \over \Re})} =1.
\label{eq14}
\ee
The overdot refers to the $d/dt$.  Solving for $\dot T$ yields
\be
\dot T ={dT \over dt} = \sqrt{ \dot \Re^2 + ({\Lambda \over 3}  \Re^2 -1 -{2M \over \Re})
\over ({\Lambda \over 3}  \Re^2 -1 -{2M \over \Re})^2},
\label{eq15}
\ee
with $\Re (t) = r_0 a(t)$.  This condition tells us the initial radius of the dust cloud
is $\Re_0=r_0 a_0$.  It also means that collapse to $a=0$ implies collapse to $\Re = 0$.

The extrinsic curvature tensor is calculated by the equation \cite{lake}
\be
K_{ij} = - n_\alpha \frac{\delta e^{\alpha}_{(i)}}{\delta \xi^j} =
-n_\alpha(\partial_j e_i^\alpha +\Gamma^\alpha_{\mu \nu} e^\mu_i e^\nu_j), \label{eq16}
\ee
where $n_\alpha$ is the unit spacelike normal to the edge, $e^{\alpha}_{(i)}$ are the basis
vectors on the edge defined by $g^{\alpha \beta}_{edge} = e^\alpha_i e^\beta_j \eta^{ij}$.
and $\xi^i$ are the coordinates on the edge.
In the interior coordinates these quantities are
\be
n_{\alpha} = \left(0,\frac{a(t)}{\sqrt{k r_0^2-1}},0,0 \right)
\label{eq17}
\ee
and
\be
e^{\alpha}_{t} = (1,0,0,0),\qquad e^{\alpha}_{\theta} = (0,0,1/r_0 a(t),0),
\qquad
e^{\alpha}_\phi = \left( 0,0,0,\frac{1}{r_0 a(t)\sinh(\theta)}\right) .
\label{eq18}
\ee
The calculations of the extrinsic curvature tensor yield
\be
K_{11}=K_{22}=\frac{\sqrt{k r_0^2 -1}}{r_0 a(t)}.
\label{eq19}
\ee
as the only non-vanishing components.

Repeating this procedure in the exterior coordinates, we find
\be
e^{\alpha}_{T} = (\dot T, \dot \Re ,0,0),\qquad e^{\alpha}_{\theta} = (0,0,1/\Re ,0),
\qquad  e^{\alpha}_\phi = \left( 0,0,0,\frac{1}{\Re \sinh \theta } \right)
\label{eq20}
\ee
for the basis vectors on the edge and
\be
n_{\alpha} = \left( - \dot \Re, \dot T, 0,0 \right)
\label{eq21}
\ee
for the unit normal. The non-zero components of the extrinsic curvature tensor are
\be
K_{00}=-\frac {d}{d\Re} \sqrt{\dot \Re^2 +\left( \frac{\Lambda}{3} \Re ^2-1-\frac {2M}{\Re }\right)
},
\label {eq22}
\ee
and
\be
K_{11}=K_{22}=\frac{\dot T}{\Re }\left( \frac{\Lambda}{3} \Re^2-1-\frac {2M}{\Re}\right) .
\label{eq22a}
\ee

The matching conditions then imply that $K_{00}$ in (\ref{eq22}) vanishes, yielding
\be
C=\dot \Re^2+\frac{\Lambda}{3} \Re^2-1-\frac {2M}{\Re},
\label{eq23}
\ee
where $C$ is simply a constant.  Using $\Re(t) = r_0 a(t)$ we find upon  matching 
(\ref{eq22a}) to (\ref{eq19}) and comparison
of (\ref{eq23}) to (\ref{eq4}) 
that $C=k r_0^2-1$ and
\be
M=\frac{1}{2} r_0^3 a \left(\dot a^2+\frac {\Lambda}{3} a^2 - k \right) =
 \frac{4}{3} \pi G \rho_0 (r_0 a_0)^3.
\label{eq25}
\ee
Hence a dust cloud of arbitrarily small
density can collapse; the boundary conditions  produce no additional constraints.
This is in contrast to $(2+1)$-dimensional collapse of dust into a black hole, which
can only proceed if the initial density is sufficiently large relative to $r_0 a_0$
\cite{ross}.  Note also that (\ref{eq23}) implies
\be
\dot \Re^2= \dot\Re_0^2 + \frac{\Lambda}{3} (\Re_0^2-\Re^2)+
{2M}(\frac {1}{\Re }-\frac{1}{\Re_0})
=r_0^2\dot{a}_0^2+ \frac {\Lambda}{3} r_0^2 (a_0^2-a^2)+\frac {2M}{r_0^2 a_0 a} (a_0-a).
\label{eq24}
\ee
which defines $\dot \Re^2$, and ensures it will always be positive for collapse provided
the constraint (\ref{eq6}) is satisfied.

As long as our conditions are satisfied, an event horzon, $R_h$, will form around
the collapsing dust. This horizon is found by setting
\be
{1 \over 3} \Lambda R_h^2-1-{2M \over R_h}=0,
\label{eq24a}
\ee
which produces one real, positive root at
\be
R_h = \frac{1}{{\Lambda}} \frac{(3M\sqrt{\Lambda}+\sqrt{9M^2\Lambda-1})^{2\over 3}+1}
{3M\sqrt{\Lambda}+\sqrt{9M^2\Lambda-1})^{1\over 3}}.
\label{eq24b}
\ee
provided $9M^2\Lambda \geq 1$.  If $9M^2\Lambda < 1$ then
\be
R_h = \sqrt{4 \over \Lambda} \cos \left({\cos^{-1}(\sqrt{9M^2 \Lambda}) \over 3} \right).
\label{eq24c}
\ee
The comoving time $t_h$ at which
collapse occurs is the time, defined by $R_h = r_0 a(t_h)$, at which the dust edge and the event horizon
are coincident, and may be found by integrating
\be
\tilde{t}_h = \int\limits_1^{x_h} dx \sqrt{ x \over {-x^3+(1+v_0^2)\left((1-B)x+B \right)}}
\label{eq25h}
\ee
where $0 < x_h \equiv a(t_h)/a_0 < 1$  since $a(t)$ is a decreasing function of $t$ and
$R_h$ is non-vanishing. The value of $\tilde{t}_h$ is finite since the integrand in (\ref{eq25h}) is
positive for $ 0 < x_h < 1$, and we necessarily have $\tilde{t}_h < \tilde{t}_c$ upon
comparing (\ref{eq9}) with (\ref{eq25h}).
 \begin{figure}
    \leavevmode
    \hfil \hbox{\epsfysize=6cm \epsffile{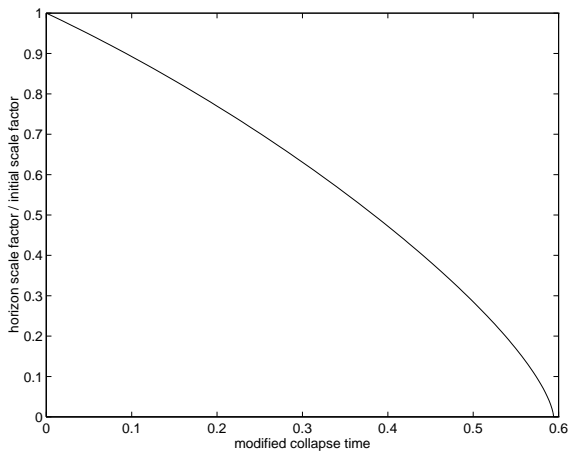}} \hfil
    \caption{A sample collapse, showing  $x_h$ shrinking as time increases.
 In this case, $v_0=1$ and $B=.5$.  This smooth collapse, with no bouncing,
guarentees that $\tilde{t}_h < \tilde{t}_c$.}
    \label{F4}
  \end{figure}

However, the coordinate time at which an external observer will witness the formation of
the event horizon is infinite, since an outgoing radial light ray emitted from the surface of
the cloud at time $T$ obeys the equation
\be
\frac{dR}{dT} = \frac{\Lambda}{3}R^2 - 1 -\frac{2M}{R}
\label{eq26}
\ee
and arrives at a point $R_F$ at time
\be
T_F = T + \int_{r_0 a(t)}^{R_F} \frac{ R dR}{\frac{\Lambda}{3}R^3 - R - 2M}
\label{eq27}
\ee
which diverges logarithmically as $r_0 a(t) \to r_0 a(t_h) = R_h$.  Hence the collapse is
unobservable from outside.  The proper time for a light source at the edge of the cloud
is equal to the comoving time, $t$, and so the emission time between wave crests of
wavelength $\lambda$ is just $dt$.  The arrival time between observed wavelengths $\lambda_F$
is just $dT_F$, and so the redshift of light from the dust edge is
\be
z \equiv \frac{\lambda_F}{\lambda} - 1 = \frac{dT_F}{dt} - 1
 = \frac{dT}{dt} - \frac{r_0 \dot{a}}{k r_0^2 -1 -r_0^2 \dot{a}^2} - 1
= \frac{1}{\sqrt{k r_0^2 -1}+ r_0 \dot{a}} - 1
\label{eq28}
\ee
which diverges as $t \to t_h$ since $r_0 \dot{a}(t_h) = - \sqrt{k r_0^2 -1}$.  Hence
the collapsing fluid fades from sight, analogous to the usual Oppenheimer-Snyder collapse
of case I.

We close this section by noting that the only curvature singularity in the spacetime
is at $R=0$. This is easily seen by computing  the Kretschmann scalar, which
in the exterior coordinates is
\be
K= 24{({\Lambda \over 3})^2 R^6 +2M^2 \over R^6}.
\ee
Alternatively, in the interior coordinates
\be
K=\frac{320(\pi G a_0^3 \rho _0)^2 - 32 \pi G a_0^3 \rho_0 a^3 \Lambda +8a^6\Lambda^2}
{3a^6}
\ee
which diverges when $t \to t_c$.

\section{Collapse of a Massless Pseudosphere}

The solution for the mass of the pseudosphere is consistent with our previous result that collapse 
requires sufficient
curvature, not density as in traditional black hole models.  This begs the question of what
will happen if the density and therefore the mass vanish. 

The interior metric is unchanged, but the exterior metric is now
\be
ds^2 = - \left( {\Lambda \over 3} R^2-1 \right)dT^2+{dR^2 \over{\Lambda \over 3} R^2-1}
+R^2(d \theta ^2 + \sinh^2 \theta d\phi^2)
\label{eq28a}
\ee
where the event horizon is $R_h=\sqrt{3 \over \Lambda}$.
The solution for the first derivative of the scale factor becomes
\be
\dot a^2 =- {\Lambda \over 3} a^2 + k.
\label{eq29}
\ee
where now
\be
k = {\Lambda \over 3} a_0^2 + \dot{a}_0^2
\ee
and the parameter $B$ is zero.  From (14) then, the `collapse' time for such a
`massless' cloud is finite, given by
\be
t_c=\sqrt{\frac {3}{\Lambda}} \int\limits_1^0 \frac {dx} {\sqrt {1+v_0^2-x^2}}
= \sqrt{\frac {3}{\Lambda}} \arcsin\left({1 \over \sqrt{1+v_0^2}}\right).
\label{eq26a}
\ee

In spite of these suggestive results, we expect that a massless cloud would be indistinguishable
from the surrounding space, and indeed we
find the interior massless metric is merely a transformation of the exterior coordinates.  The
transformation is:
\be
R=r \cos\left(\sqrt{\Lambda \over 3} t \right) , \qquad
 \sinh \left( \sqrt{\Lambda \over 3} T \right)
={\sin(\sqrt{\Lambda \over 3}t) \over \sqrt{{\Lambda \over 3}[r \cos(\sqrt{\Lambda \over
3}t)]^2-1}},
\ee
\be
\hat \theta = \theta, \qquad \hat \phi = \phi.
\ee
Here, $k = {\Lambda \over 3}$ and $a$ has been transformed to a cosine function.
The `collapse' time has become simply the time at which the identification surfaces used to
construct the $(3+1)$-dimensional AdS black hole (\ref{eq28a}) merge \cite{amin}.  The curvature is no
longer singular at $R=0$, and the singularity is of the Misner type.

\section{Conclusions}

We have demonstrated that a cloud of pressureless dust can undergo gravitational collapse
to a black hole whose event horizon has arbitrary topology.  The properties of the collapsing
dust are similar to those of the usual (case 1) Oppenheimer-Snyder collapse, with black holes
of arbitarily small mass being formed from arbitarily small initial density distributions.
Perhaps the most unusual feature of the collapse scenario investigated here is that,
for a given value of the cosmological constant,
collapse is not possible once the density is sufficiently large unless the dust cloud
is given a large enough initial inward velocity.  This feature is a consequence of the
negative curvature of the $(\theta, \phi)$ sector in the spacetimes we consider. 

Of course the exterior black hole spacetimes to which the dust collapses have non-trivial
topology.  By enclosing the entire dust cloud and a portion of the vacuum inside a
shell of appropriate stress-energy and topology, it might be possible to match these
solutions onto background spacetimes whose topology is ${\bf R}^2\otimes{\bf S}^2$, using
methods similar to those discussed in refs. \cite{aur,lake}.  Work
on this is in progress.

\section{Acknowledgments}
This work was supported in part by the Natural Sciences and Engineering Research
Council of Canada.

\end{document}